# Compound Attention and Neighbor Matching Network for Multi-contrast MRI Super-resolution

Wenxuan Chen, Sirui Wu, Shuai Wang, Zhongsen Li, Jia Yang
Huifeng Yao, and Xiaolei Song

**Abstract—Multi-contrast magnetic resonance imaging (MRI) reflects information about human tissue from different perspectives and has many clinical applications. Multi-contrast super-resolution (SR) of MRI could synthesize high-resolution images of one modality from the acquired low-resolution (LR) images, by utilizing the complementary information from another modality. An important prior knowledge of multi-contrast MRI is that, multi-contrast images are obtained with similar or same field-of-views (FOVs). However, existing methods did not exploit this prior; only performing simple concatenation of the reference and LR features, or global feature-matching. Herein, we proposed a novel network architecture with compound-attention and neighbor matching (CANM-Net) for multi-contrast MRI SR. Specifically, to effectively utilize the priors of similar FOVs, CANM-Net proposed a neighborhood-based feature-matching method, which only calculates the similarity of a LR patch with the corresponding patch on the HR reference and its adjacent ones; the compound self-attention with a pyramid-structure effectively captures the dependencies in both spatial and channel dimension. We conduct experiments on the IXI, fastMRI, and in-house datasets, with T2-, T1-weighted images as the LR and reference images, respectively. In 4× and 2× SR tasks, the CANM-Net outperforms state-of-the-art approaches in both retrospective and prospective experiments. The ablation study proves the rationality of CANM-net. Additionally, when the input LR images and the reference HR are imperfectly registered, CANM-Net still achieves the best performance among all test methods. In summary, CANM-Net may have potential in clinical applications, for recovery and enhancement of multi-contrast MRI.**

## I. INTRODUCTION

MAGNETIC resonance imaging (MRI) is a non-invasive medical imaging technology that has gradually been widely applied in hospitals, because it provides more information on the soft tissue while avoiding ionizing radiation [1, 2]. However, some physical defects in imaging systems inevitably result in difficulties in the acquisition of high-resolution (HR) MR images [3-5]. Most commonly, the long scanning time of HR images is a huge burden for practical applications, which not only causes patient discomfort, but may also lead to serious motion artifacts for the abdomen and chest imaging [6, 7]. Besides, the low signal-to-noise ratio (SNR) is another limitation of the acquisition of HR images. The super-resolution (SR) reconstruction of MR images is a promising way to provide MR images with high resolution without updating hardware facilities [8].

Multi-contrast MRI can provide complementary information about the same region from various aspects [9-11]. For example, T1-weighted images (T1WIs) can provide images with clear anatomic structural information in most situations, while T2-weighted images (T2WIs) are usually more effective in detecting lesions such as ischemia and hemorrhage. Importantly, MR images in different modalities have often different physical characteristics due to the varied acquisition procedures. For example, high-resolution (HR) T1WIs are relatively faster to acquire than T2-weighted images (T2WIs), because T1WIs require shorter echo time (TE) and repetition time (TR). Similarly, the acquisition of HR proton-density-weighted images (PDWIs) is more rapid than that of fat-suppressed proton-density-weighted images (FS-PDWIs) [12]. In most clinical situations, people simultaneously acquire MR images in different modalities to have better diagnoses of diseases. Therefore, how to utilize the information in the easy-to-acquire modality as the reference to guide the reconstruction of images in the hard-to-acquire modality (i.e., requiring longer acquisition time), has become a research focus recently.

Many traditional methods for the acceleration of MR images based on a single modality, such as parallel imaging methods [13-15], compressed sensing (CS) [16], iterative-based algorithms [17-19], dictionary learning [20], and low rank [21], have been extensively studied, and they have both strengths and weaknesses compared with deep learning (DL) based methods. Multi-contrast-based methods can usually achieve better performance than those based on a single image, since the auxiliary information is an effective guidance for the reconstruction of hard-to-acquire images [22-24]. Recently,

This work was supported by National Nature Science Foundation of China (820719114).

W. Chen, S. Wu, S. Wang, Z. Li, and X. Song are with the Center for Biomedical Imaging Research, Tsinghua University (Beijing), 100084, China. (Email: cwx21@mails.tsinghua.edu.cn; wsr21@mails.tsinghua.edu.cn; bit.ybws@gmail.com; lizhongsen21@mails.tsinghua.edu.cn; songxl@tsinghua.edu.cn.) J. Yang is with the School of Materials Science and Engineering, Tsinghua University (Beijing), 100084, China.

H. Yao is with the Department of Computer Science and Engineering, Hongkong University of Science and Technology. (Email: hyaoad@connect.ust.hk.)

Corresponding author: Xiaolei Song.

DL-based approaches have the tendency to become mainstream for multi-contrast methods because of their superior ability to fuse complementary semantics in different modalities [25-27]. More specifically, most multi-contrast-based MRI SR approaches are faced with two key challenges: (a) how to effectively capture the feature in both reference and LR images, and (b) how to match the corresponding feature between reference and LR images and fuse them. In the earlier studies, Lyu et al. [28] proposed a convolutional neural network (CNN) architecture with two branches receiving the reference and LR images, employing the channel-dimension concatenating at the bottleneck for the feature fusion. Feng et al. [29] proposed a multi-stage integration architecture to fuse the feature, and they [30] later proposed a separable attention mechanism for feature extraction in their subsequent work. However, all of the above methods are based on CNN-only networks that are deficient in extracting non-local features. More recently, in the work of Li et al. [31] and Feng et al. [32], they employ transformer [33] layers to capture the long-range dependencies in the images. Moreover, [31] introduced the feature matching mechanism, which was earlier proposed in the reference-based super-resolution (Ref-SR) tasks in the natural image domain, into the multi-contrast MRI SR and achieved good performance.

However, existing methods still have deficiencies in their design of network architecture that limit their performance. There are two main shortcomings as follows: First, some methods including [31, 34] apply a coarse-to-fine global feature matching between the features of reference and LR images; however, global attention that achieves good performance in Ref-SR tasks may not be unsuitable for multi-contrast MRI SR, which possibly causes increased computational complexity and possible overfitting. (We will detailly discuss the distinction between Ref-SR and multi-contrast MRI SR in **Section II B.**) Second, although spatial attention and the use of shifted-windows strategy have been utilized in some existing methods [30-32] utilize, they are still not ideally suited to low-level vision tasks including super-resolution.

To address these shortcomings, we propose a compound-attention and neighbor-matching network (CANM-Net) for multi-contrast MRI SR. Taking both reference and LR images as inputs, our method employs compound transformer layers that consist of window-attention and channel-attention to effectively capture the dependencies in both spatial and channel dimensions. Then, neighborhood-based feature matching modules are applied to match the features of reference and LR images in a multi-scale form. The matched features are then fused, and undergo upsampling in the decoder, finally providing the restored outputs with high quality. To investigate the effectiveness of the proposed CANM-Net, we conduct extensive experiments of multi-contrast MRI SR task on three MRI datasets: IXI (brain; T1WI and T2WI) [35], fastMRI (knee; PDWIs and FS-PDWIs) [36], and real-world in-house MRI (brain; T1WI and T2WI; prospective study) datasets. Overall, our main contributions can be summarized as follows:

(a) We proposed a novel network architecture with compound attention and neighbor matching for multi-contrast

MRI SR. We discussed the similarities and differences between Ref-SR and multi-contrast MRI SR, and proposed the neighbor-based feature matching scheme that can effectively match and fuse the features from reference and LR images.

(b) The compound attention mechanism, which exists in both the encoder and decoder, captures the dependencies in the spatial and channel dimensions simultaneously. In addition, we improve the channel-based attention to have a pyramid structure, which enhances the robustness of the network.

(c) Experimental results on three datasets show the CANM-Net outperforms state-of-the-art methods in both qualitative and quantitative evaluation. The robustness study proves that the CANM-Net can still achieve good performance when the reference and LR images are imperfectly registered and have different orientations. These results demonstrate the CANM-Net's effectiveness and good potential in practical applications. The code of our study is available at https://github.com/ChenWenxuan2021/CANM_Net.

## II. RELATED WORKS

### A. Single Image Super-resolution in MRI

Image super-resolution models take the low-resolution (LR) images as inputs to provide HR images, which are usually highly ill-posed. Before the rise of DL, many traditional algorithms had been proposed to solve single MR image super-resolution tasks: parallel-imaging-based methods [13-15], low-rank [21], dictionary learning [20], iterative-based [17-19], and prior-based methods [37, 38] in existing studies are proposed for the SR task of MRI images. However, when faced with more challenging tasks, such as SR reconstruction with upscale factors of larger than 4, these traditional methods usually cannot obtain outputs with satisfactory quality. Recently, DL has received much attention and has been widely used in many low-level vision tasks including image SR. DL-based methods utilize large-scale datasets for training, extracting the inherent feature of the image, and usually achieve better performance than traditional methods. Besides, with the less requirement of the image prior information, those DL-based methods proposed for X-ray, computed tomography (CT), and even natural images can also be applied for MRI SR [39-43]. Meanwhile, researchers also make efforts to design MRI SR methods based on the characteristics of MR images [44-47]. However, these methods only utilize the information of images in a single modality.

### B. Multi-contrast SR in MRI and Ref-SR for Natural Images

Compared to single-image SR methods, multi-contrast MRI SR utilizes the auxiliary information in the reference to guide the reconstruction of LR images, usually obtaining outputs with higher quality [48, 49]. Earlier studies [28-30] usually fused the reference and LR features by directly connecting them. Li et al. [31] notice that both multi-contrast MRI SR and Ref-SR have the concern to transfer the texture information from the reference to LR images. Inspired by Ref-SR methods, they introduced feature matching into their methods for multi-

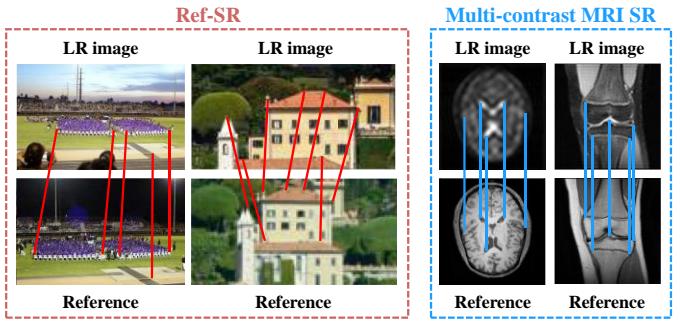

**Ref-SR**      **Multi-contrast MRI SR**

Fig. 1 **The comparison between Ref-SR and multi-contrast MRI SR tasks.** LR and reference images in the Ref-SR tasks are usually obtained from varied views, whereas MR images in different modalities are often in the similar FOVs.

contrast MRI SR, and their method outperformed previous methods which directly concatenated the feature maps. However, they utilize a global matching scheme in their work, which can be further improved to be more suitable for multi-contrast MRI SR. Here we discuss the difference between Ref-SR and multi-contrast MRI SR, as shown in **Fig. 1**: In most situations, the LR and reference images in the Ref-SR tasks are often obtained from varied views [50, 51], whereas they usually have the similar field-of-views (FOVs) in the multi-contrast MRI SR tasks. Even though MR images in different modalities were acquired in different FOVs (i.e., different orientations), one can apply image registration to effectively align them [52]. This prior information reminds us to propose a neighborhood-based feature matching mechanism between varied modalities.

### C. Self-attentions in Vision Transformers

While convolutional layers have limited reception fields and cannot adequately capture the long-range dependencies in the feature, the vision transformer and its follow-up work utilize the self-attention mechanism and can extract global features in images [54, 56]. Some existing methods have employed transformers in their feature extraction modules and shown good effectiveness. Feng et al. proposed a task transformer to combine the reconstruction and SR of MR images [45], and they also introduced a multi-modal fusion transformer to accelerate multi-contrast MR imaging [32]. Li et al. [31] utilized the shifted-window transformer (swin-transformer) [56] groups in their work to combine the advantages of CNN and transformers. To obtain a better understanding of self-attention operations in transformers, we roughly classify transformers into the following three categories.

(a) Vanilla-ViT-like networks [54, 55]: They embed the image into N unfolded patches, and the self-attention is performed among patches. These methods suffer from huge computational complexity when input's resolution is high.

(b) Swin-transformer-like networks [56]: The widely-known swin-transformer first partitions features into windows, and the self-attention operation is conducted inner each window. A shifted-window strategy is introduced to help the pixels at the window edge share context information with pixels in other windows.

(c) Restormer [57]: Zamir et al. proposed a transformer-based image restoration network named restormer, which first utilizes three convolutional layers, instead of fully-connected layers, to obtain the Q, K, and V. Then, the Q, K, and V are *flattened* to a shape of (C, H×W), and the self-attention operation is calculated to obtain the (C×C)-shape attention matrix among the channels. Restormer has shown good effectiveness in various low-level image restoration tasks including deraining, deblurring, denoising, etc.

In our method, we combine the window-attention and channel-attention in both the encoder and decoder and apply a pyramid structure to enhance the robustness of the network.

## III. METHODS

Overall, the proposed CANM-Net has a U-shape structure as shown in **Fig. 2**. The encoder has two branches to respectively receive the reference and LR images, each consisting of four compound transformer encoding (CTE) modules. Unlike most existing methods that only perform feature matching in the bottleneck layer, our method has three independent neighborhood-based feature matching (NBFM) modules for feature fusion at different scales. After the feature matching and texture transferring, three compound transformer decoding (CTD) modules are utilized to sequentially upsample the fused feature, finally obtaining the HR outputs.

### A. Compound Transformer: Encoding and Decoding

As mentioned above, long-range dependencies and feature interactions in both the spatial and channel dimensions are important for subsequent feature matching and texture transfer. Therefore, we propose the CTE and CTD modules to effectively extract the features. CTE and CTD have almost the same architecture: as shown in **Fig. 3**, they are composed of several compound transformer layers (CTL), which further contain a window-attention block (WAB), a channel-attention block (CAB), and a feed-forward block (FFB). The operations in the WABs are similar to those in the swin-transformer. Feature maps are firstly partitioned into small windows, and the self-attention operations are performed inner each window. Shifted-window strategy for pixels to exchange information with their neighbor windows is also applied to the WABs in the (2nd, 4th, …, Nth) CTLs, where N is the number of CTLs and is even.

The CABs in our method are improved based on the original channel attention in the restormer. We noticed that the original self-attention operations are performed with pixel-wise multiplication, which means a quadratic relationship between the computational complexity and the feature map scale, causing a high calculation cost for the low-level features with relatively large scales. Moreover, pixel-wise multiplication is sensitive to the change in the input's relative position. The result of matrix multiplication may dramatically change when the relative positions between the reference and LR images have a slight shift. Considering these deficiencies, we introduce the pyramid structure into the CABs of our method to reduce the calculation cost and enhance the robustness. As shown in

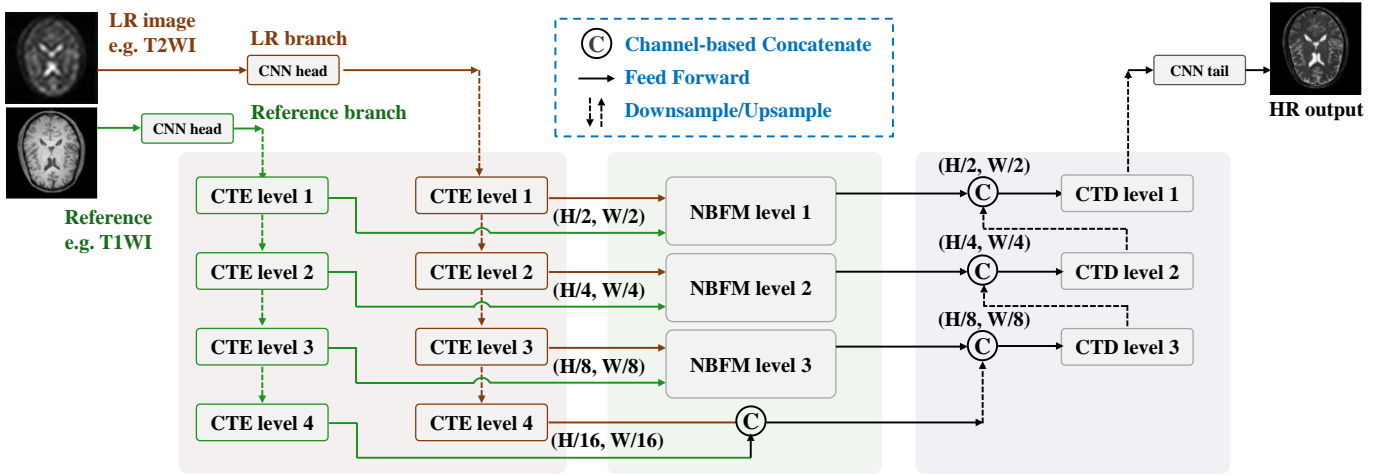

Fig. 2 **The overall network architecture of the CANM-Net.** Two branches, each with a CNN head and four CTE modules, are utilized to receive the reference and LR images, and their features are matched and fused in the NBFM modules. The decoder has skip connections similar to the U-Net, finally obtaining the restored outputs.

**Fig. 3**, the $V$ of the self-attention is still obtained from features with the original scale. However, we first downsample the features with factors of 2 and 4 before obtaining $Q$ and $K$. The attention operation at the $1/2\times$ and $1/4\times$ scale will only require 1/4 and 1/16 times of calculation compared to the operation at the original scale. Then, we added these two (C×C) matrixes with two learnable parameters $\alpha_1$, $\alpha_2$ as their weight. Finally, the mixed attention matrix undergoes the softmax and provides the outputs by multiplying the $V$. The expressions of WABs and CABs can be described as

$$WAB(x) = Sfm\left(\frac{toQ(x) * toK(x)^T}{\sqrt{d}}\right) toV(x) \tag{1}$$

$$attn_{/2} = toQ_{/2}(ds_{/2}(x)) * toK_{/2}(ds_{/2}(x))^T \tag{2}$$

$$attn_{/4} = toQ_{/4}(ds_{/4}(x)) * toK_{/4}(ds_{/4}(x))^T \tag{3}$$

$$CAB(x) = Sfm\left(\frac{\alpha_1 * attn_{/2} + \alpha_2 * attn_{/4}}{\sqrt{d}}\right) toV(x) \tag{4}$$

where $toQ$, $toK$, and $toV$ represent the converting from features to the $Q$, $K$, and $V$, $Sfm$ represents the softmax function, $ds$ represents the downsampler, and $\sqrt{d}$ is the dimensional scale factor in the transformers. The outputs of WAB and CAB are then concatenated and go through a dimensional-reduction layer. Finally, we employ the FFB, which consists of depth-wise convolution with the gating mechanism instead of fully-connected layers, to conduct the interaction among channels of features. Residual connections are applied to all the CTLs to enhance stability during the network training. Overall, we have the CTLs as

$$CTL(x) = x + FFB\left(dr(concat(WAB(x), CAB(x)))\right) \tag{5}$$

where $dr$ and $concat$ are the dimensional-reduction and concatenate. At the end of each CTE module, a downsampler is exploited to reduce the size of features, whereas upsamplers are applied at the end of CTD modules. Besides, the skip connection similar to the U-Net is used for each CTD module to receive outputs from both the previous CTD and the encoder.

### B. Neighborhood-based Feature Matching

The feature fusion and texture transfer between reference and LR images also determine the quality of outputs in both Ref-SR and multi-contrast MRI SR. Feature matching schemes in previous methods can be roughly divided into the following two categories: (a) one-step global matching, in which each patch in reference features is required to calculate similarity with all patches in LR features, and (b) coarse-to-fine matching, where features are divided into several blocks (the size of blocks ≫ patches), with matching the blocks in the first step. In the second step, patch matching is conducted inner the corresponding reference and LR blocks that have been matched in the first step. Although coarse-to-fine matching requires less calculation than one-step global matching, it still belongs to the global matching mechanism. As discussed in previous sections, global matching has a high calculation cost and may lead to overfitting that affects its performance in the multi-contrast MRI SR.

Therefore, we introduce the NBFM modules into the CANM-Net to match the features from reference and LR images. Before the main feature of NBFM, adaptive instance normalization (AdaIN) [53] is applied to the reference features to avoid the influence of data distribution differences among the reference and LR images. AdaIN firstly converts the intrinsic deviation and mean of reference features into 1 and 0 as

$$IN_{no\_bias}(x_{ref}) = \frac{x_{ref} - \mu(x_{ref})}{\sigma(x_{ref})} \tag{6}$$

where the $\mu$ and $\sigma$ represent the operation of calculating mean and standard deviation. Then, two convolutional layers are independently utilized to extract the $\gamma$ and $\beta$ from the LR features, which are then exploited to control the style of reference features:

$$\gamma = Conv_{gamma}(x_{deg}), \beta = Conv_{beta}(x_{deg}) \tag{7}$$

$$AdaIN(x_{ref}) = \gamma * IN_{no\_bias}(x_{ref}) + \beta \tag{8}$$

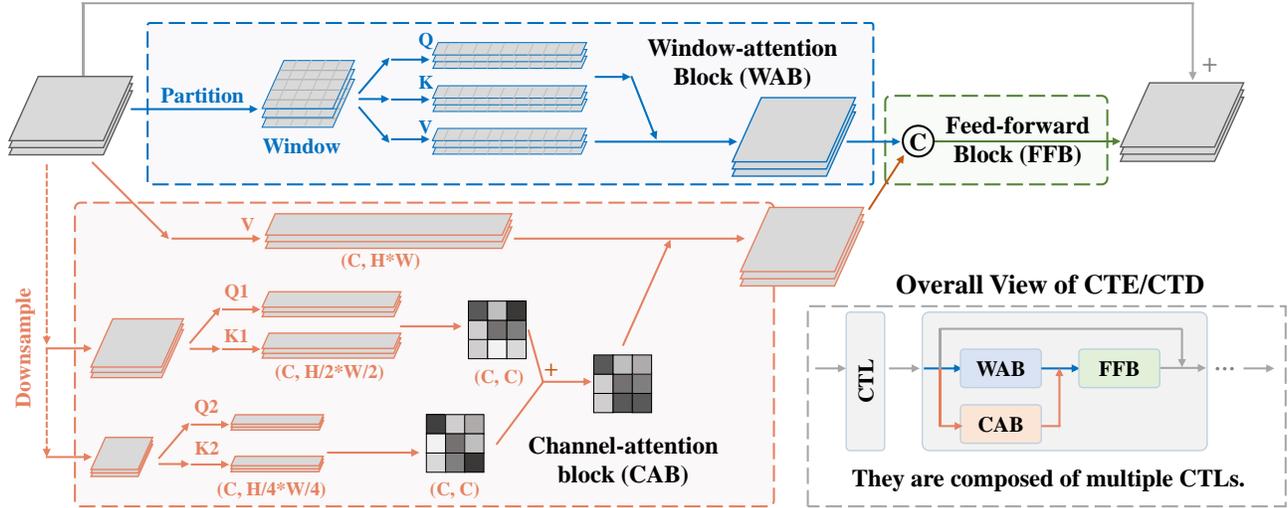

Fig.3 **The Network architecture of compound transformer encoders and decoders** is composed of a window-attention block, a channel-attention block, and a feed-forward block, and can capture the feature interactions in both the spatial and channel dimension. To enhance the stability and reduce the calculation cost, a pyramid structure of Q and K is employed in the CAB.

The main architecture of NBFM is shown in **Fig. 4**, where the $x_{ref}$ has already undergone the AdaIN operation. We partition both the reference and LR features into M patches. For each LR patch, the cosine similarity is calculated with all its surrounding reference patches. This processing can be described as

$$S(i,j) = < p_{deg}(i), p_{ref}(j) > \tag{9}$$

where $p$ represents patches; $i$ represents the $i^{th}$ LR patch in LR features; $j$ is the $j^{th}$ reference patch in the neighborhood of $p_{deg}$; and $<\cdot>$ represents the calculation of cosine similarity. Obviously, we have the $i \gg j$, thus reducing the computational complexity. After obtaining the similarity matrix, we conducted element-wise multiplication with a learnable parameter matrix that has the same shape. Then, instead of choosing the patch with the highest similarity score, we utilize the soft-attention method to integrate all semantics of reference features in the neighborhood. As shown in **Fig. 4**, the attention matrix first goes through a softmax operation to have a sum of 1. Then, we fuse all the reference patches in the neighborhood by adding them together, with the attention matrix as their weights. The above-mentioned operations can be formulated as

$$attn(i,j) = sfm(S(i,j) * W) \tag{10}$$

$$p_{ref}^m(i) = \sum_{j}^{N} attn(i,j) * p_{ref}(j) \tag{11}$$

where $S$, $W$, and $attn$ are the similarity matrix, learnable weights, and the attention matrix, respectively; $N$ is the number of reference patches in the neighborhood. Finally, we fold the patches back to feature maps with their original shape, and concatenate the matched reference features with the LR features to fuse them, employing a convolutional layer with the $1\times1$ kernel size to reduce the dimension of outputs.

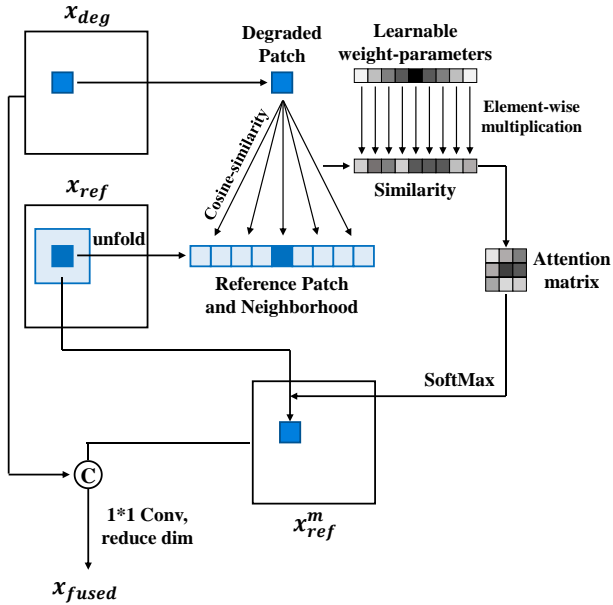

Fig. 4 **Main architecture of the neighborhood-based feature matching module.** Reference and LR features are first unfolded into patches. Then, the LR patches only calculate the similarity with corresponding patches and their neighboring patches in the reference features. A Learnable weight-parameter matrix is employed to control the matching in an overall manner.

## IV. EXPERIMENTS

### A. Experimental Details

#### 1) Datasets:
All the experiments in this paper are conducted on the following three MRI datasets:

(a) **IXI dataset** [35] is an open-access MRI dataset for the brain images, officially provided in the http://brain-development.org/ixi-dataset/. T1WIs and T2WIs of IXI dataset are utilized in our work. Specifically, we first registered the T1- and T2-volumes in 3D form. We randomly selected 282, 81, and 39 pairs of registered volumes for training, testing, and validation. Then, each T1 and T2 volume was cut into 2D images in the cephalo-caudal direction with a slice thickness of 2 mm. We only preserved the central 72 slices of brain volumes to remove the skull-only or contentless images. Finally, there are 20304, 5832, and 2808 pairs of T1WIs and T2WIs in the training, testing, and validation set, each with a spatial resolution of 256×256. T1WIs are regarded as the reference to guide the reconstruction of T2WIs.

(b) **FastMRI dataset** [36] is the largest open-access MRI dataset that is available at https://fastmri.med.nyu.edu/. We chose the PD-weighted and FS-PD-weighted knee images of fastMRI in the experiments. Similarly, there were a total of 584, 168, and 80 pairs of volumes, cutting into 18688, 5376, and 2560 pairs of 2D images for training, testing, and validation, respectively. PDWIs are the reference images, and FS-PDWIs are the targets of super-resolution.

(c) **An in-house acquired MRI dataset** of brain T1WIs and T2WIs was acquired on the Philips Ingenia CX 3T clinical scanner. The MR acquisition procedures are close to those of IXI dataset, and are mainly as follows: for T1WIs, TE = 4.8 ms, TR = 10 ms; for T2WIs, TE = 110 ms, TR = 5000 ms; both the T1WIs and fully-sampled T2WIs have the field of view (FOV) as 240×240 mm, and the resolution as 256×256. Besides, 1/4× and 1/2× LR T2WIs were also acquired with the resolution of 64×64 and 128×128, respectively. We totally scanned 24 cases of healthy volunteers, in which five volumes were used as the testing set.

*2) Data preparation:* We conducted both retrospective and prospective studies to evaluate the performance of our method in the multi-contrast SR task, and their data preparation pipelines are as follows:

(a) **Retrospective study**: We conducted the retrospective study of the **SR** task with both the upscale factor of 4 and 2 (shorted as 4× SR and 2× SR). First, we converted multi-contrast images into k-space, only preserving the (1/4, 1/4) and (1/2, 1/2) central regions of the k-space to reduce the resolution.

(b) **Prospective study**: The prospective study was conducted only on the in-house dataset acquired on the 3T clinical scanner. The real-acquired LR images together with the reference images are directly utilized as inputs.

*3) Model Implementation:* As mentioned in **Section III** and shown in **Fig. 2**, the CANM-Net has eight CTE modules (4×2 for the two branches), three NBFM modules, and three CTD modules. Other hyperparameters are shown in **Table 1**.

Table I: Hyperparameters of the CANM-Net and its training strategy.

| Name | Value |
|---|---|
| Input resolution | (256, 256), and interpolation is required for the LR images in SR task |
| Feature-map shape | from (128, 128) to (16, 16) in the 1st to 4th-level features |
| Number of channels | from 64 to 256 in the 1st to 4th level |
| Number of CTLs | [4, 4, 16, 4] in the 1st to 4th-level CTEs, and [4, 4, 4] in the 1st to 3rd-level CTDs |
| Number of WA heads | [2, 4, 8, 8] in the 1st to 4th-level CTLs |
| Number of CA heads | [1, 2, 4, 8] in the 1st to 4th-level CTLs |
| Neighborhood size | [(3, 3), (3, 3), (5, 5)] in the 1st to 3rd-level NBFM modules |
| Patch size | (3, 3) in all NBFM modules |
| Batch size | 4 |
| Learning rate | begin at 1*10⁻⁴, with an attenuation of 0.988 |
| Maximum of epoch | 200 |

The codes were implemented by PyTorch with four NVIDIA RTX 3090 GPUs and 24 GB memory in each GPU. Since the real-world in-house dataset had a much smaller scale than the

IXI and fastMRI datasets, its training started from the pre-trained weighted on the IXI dataset. Following the work of [32, 48], we only use the L1-loss to constrain the network, with the form as

$$L_1(Y, T) = \|Y - T\|_1 \quad (12)$$

where $Y$ and $T$ are the outputs and target images.

*4) Baselines:* We compared our method with the following four groups of state-of-the-art methods.

(a) **Modality transfer methods** comprise pix2pix [58] and PTNet [59].

(b) **Widely-used transformer-based methods** comprise *swinIR* [60] and restormer [57]. They have shown good performance in many low-level vision tasks including SR in previous studies.

(c) **Ref-SR methods** comprise TTSR [51], CrossNet++ [61], MASA [32], and DATSR [62]. Methods in this group were first introduced for the SR task of natural images, mainly focusing on the feature matching mechanism between LR and reference images.

(d) **Existing multi-contrast MRI SR methods** comprise MINet [29], SANet [30], and McMRSR [31].

The methods in groups (a) and (b) only require one input, and therefore we concatenate the reference and LR images (with zero-padding in the k-space) into a two-channel image to have a fair comparison. TTSR, MASA, and DATSR were not evaluated in the 2× SR task because they cannot be performed without changing their network architectures.

*5) Quantitative Evaluation:* To quantitatively compare the performance of our methods and baselines, we evaluate the experimental results on the testing set with the widely-used metric peak signal-to-noise ratio (PSNR) and structural similarity (SSIM). Larger PSNR and SSIM mean better results.

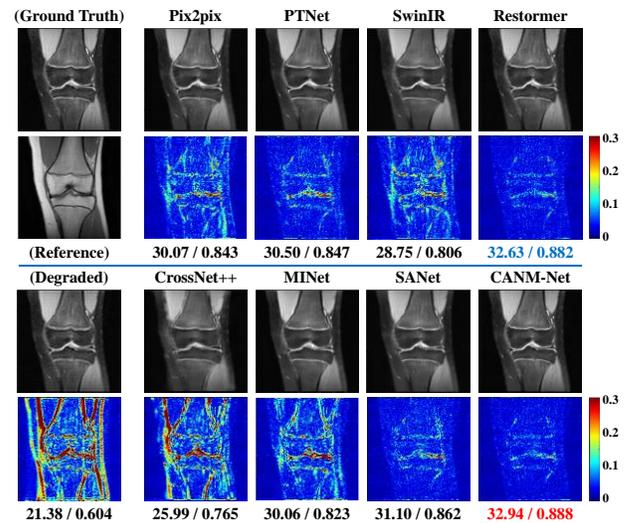

Fig. 5 **Qualitative comparison of the 4× SR task on the fastMRI dataset.** Both the restored images and their error maps with the ground truth are presented. Higher values in the error map mean a worse result.

### B. Comparative Results

*1) Results on IXI and fastMRI datasets:* The results of the retrospective experiment conducted on the IXI and fastMRI

datasets are shown in **Table 2**, where we have divided these baseline methods into four groups based on their mechanisms. It can be seen that the transformer-based image restoration methods (swinIR and restormer) provide relatively good results in most tasks. Interestingly, McMRSR outperforms MASA in all experiments although they only have the difference of whether to employ swin-transformers, suggesting the essential to extract global features in the SR task. Restormer provides the second-best results in all experiments involved in **Table 2**, proving the effectiveness of channel-attention in the low-level vision tasks. As shown in the last row of **Table 2**, the proposed CANM-Net has the best performance on the IXI and fastMRI datasets in both the tasks of 4× SR and 2× SR.

To further study whether the CANM-Net has a significant superiority, we conduct statistical analysis and student's T-test between the CANM-Net and other baseline methods, regarding the p-value of 0.001 as the boundary. Our method achieves significantly better performance than almost all baselines except for the restormer of the 4×SR task on the IXI dataset.

We select a representative FS-PD weighted knee-image of

4× SR task on the fastMRI dataset for the qualitative comparison of the CANM-Net and baselines. The results are shown in **Fig. 5**: the 1st and 3rd rows are the SR outputs, and the 2nd and 4th rows show the error maps between outputs and ground truth. As can be observed, our method yields the output with the best overall visual effect and the least error to ground truth, suggesting an improved performance than baseline methods.

*2) Results on the in-house Dataset:* Both retrospective and prospective experiments are conducted on the in-house MRI dataset, with real-acquired HR T1WIs and LR T2WIs as reference and LR images of the prospective study, and HR T2WIs as targets in the loss-function. Experimental results are shown in **Table 3**. As can be observed, all methods in the prospective study have a decrease in both PSNR and SSIM compared to the retrospective study, which may be due to the prospective image super-resolution being more challenging than retrospective tasks. However, our method still has the best result compared to baselines, which suggests a good potential in clinical applications. The statistical results of PSNR in the prospective study of the 4× SR tasks, is shown in **Fig. 6**, together with those on IXI and FastMRI datasets. The CANM-Net significantly outperforms state-of-the-art methods in the prospective study of the in-house dataset.

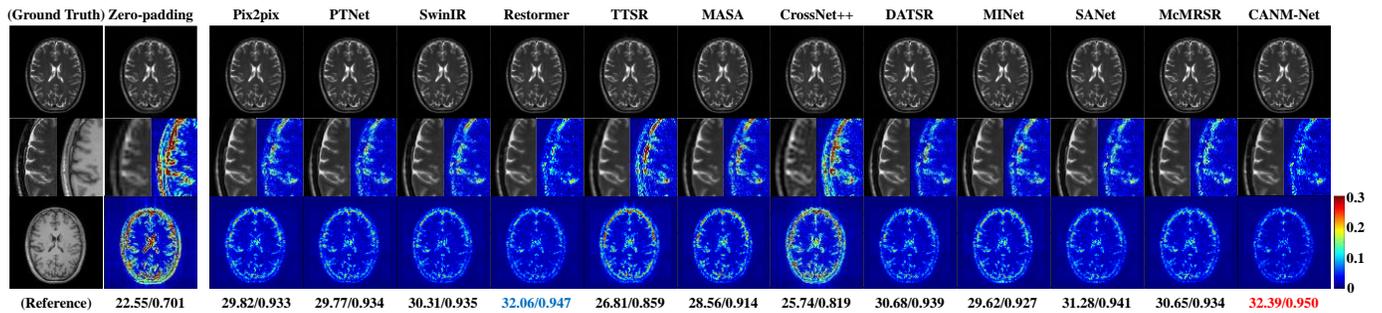

Fig. 6 **Qualitative comparison of the 4× SR task in the prospective study of the in-house dataset.** The upper, lower, and middle rows are outputs, error maps, and the enlargement of detailed structures, respectively. Higher values in the error map mean a worse result.

Table II: Retrospective experiments on the IXI and fastMRI dataset (PSNR and SSIM). The best and second-best results are in **red** and **blue**.

| IXI and FastMRI dataset | | IXI, 4× SR | | | IXI, 2× SR | | | FastMRI, 4× SR | | | FastMRI, 2× SR | | |
|---|---|---|---|---|---|---|---|---|---|---|---|---|---|
| | | PSNR | SSIM | p value | PSNR | SSIM | p value | PSNR | SSIM | p value | PSNR | SSIM | p value |
| Modality transfer methods | Pix2Pix [58] | 31.62 | 0.953 | <0.001 | 37.88 | 0.983 | <0.001 | 29.37 | 0.845 | <0.001 | 34.64 | 0.972 | <0.001 |
| | PTNet [59] | 32.44 | 0.961 | <0.001 | 38.65 | 0.989 | <0.001 | 30.06 | 0.866 | <0.001 | 35.55 | 0.975 | <0.001 |
| Transformer-based restoration methods | SwinIR [60] | 33.58 | 0.969 | <0.001 | 38.80 | 0.989 | <0.001 | 31.18 | 0.879 | <0.001 | 36.90 | 0.981 | <0.001 |
| | Restormer [57] | **34.04** | **0.973** | 0.003 | **39.85** | **0.991** | <0.001 | **31.79** | **0.885** | <0.001 | **37.62** | **0.985** | <0.001 |
| Ref-SR methods | TTSR [51] | 29.50 | 0.868 | <0.001 | | | | 27.62 | 0.844 | <0.001 | | | |
| | MASA [32] | 32.24 | 0.957 | <0.001 | | | | 30.75 | 0.879 | <0.001 | | | |
| | CrossNet++ [61] | 30.84 | 0.946 | <0.001 | 38.03 | 0.985 | <0.001 | 29.13 | 0.840 | <0.001 | 35.90 | 0.973 | <0.001 |
| | DATSR [55] | 32.88 | 0.964 | <0.001 | | | | 31.07 | 0.884 | <0.001 | | | |
| Existing multi-contrast MRI SR methods | MINet [29] | 32.30 | 0.959 | <0.001 | 38.44 | 0.985 | <0.001 | 30.21 | 0.871 | <0.001 | 35.22 | 0.972 | <0.001 |
| | SANet [30] | 33.30 | 0.966 | <0.001 | 39.58 | 0.988 | <0.001 | 30.75 | 0.881 | <0.001 | 36.49 | 0.976 | <0.001 |
| | McMRSR [31] | 33.16 | 0.964 | <0.001 | 38.68 | 0.988 | <0.001 | 31.40 | 0.880 | <0.001 | 36.65 | 0.978 | <0.001 |
| The proposed | CANM-Net (Ours) | **34.44** | **0.974** | —— | **40.65** | **0.993** | —— | **32.45** | **0.897** | —— | **38.01** | **0.986** | —— |

Table III: Retrospective and prospective experiments conducted on the in-house MRI dataset. The best and second-best results are in **red** and **blue**.

| In-house dataset | | Retrospective, 4× SR | | | Retrospective, 2× SR | | | Prospective, 4× SR | | | Prospective, 2× SR | | |
|---|---|---|---|---|---|---|---|---|---|---|---|---|---|
| | | PSNR | SSIM | p value | PSNR | SSIM | p value | PSNR | SSIM | p value | PSNR | SSIM | p value |
| Modality transfer methods | Pix2Pix [58] | 30.92 | 0.947 | <0.001 | 35.15 | 0.962 | <0.001 | 28.98 | 0.922 | <0.001 | 33.22 | 0.952 | <0.001 |
| | PTNet [69] | 30.85 | 0.940 | <0.001 | 35.32 | 0.968 | <0.001 | 29.07 | 0.927 | <0.001 | 33.58 | 0.960 | <0.001 |
| Transformer-based restoration methods | SwinIR [60] | 31.03 | 0.934 | <0.001 | 35.27 | 0.947 | <0.001 | 28.48 | 0.908 | <0.001 | 33.76 | 0.959 | <0.001 |
| | Restormer [57] | **32.55** | **0.953** | <0.001 | **37.47** | **0.975** | <0.001 | **30.86** | **0.943** | <0.001 | **35.53** | **0.969** | <0.001 |
| Ref-SR methods | TTSR [51] | 27.40 | 0.845 | <0.001 | | | | 27.44 | 0.887 | <0.001 | | | |
| | MASA [32] | 29.35 | 0.904 | <0.001 | | | | 28.59 | 0.904 | <0.001 | | | |
| | CrossNet++ [61] | 25.78 | 0.797 | <0.001 | 34.61 | 0.944 | <0.001 | 27.19 | 0.881 | <0.001 | 32.15 | 0.927 | <0.001 |
| | DATSR [55] | 31.10 | 0.935 | <0.001 | | | | 29.65 | 0.909 | <0.001 | | | |
| Existing multi-contrast MRI SR methods | MINet [29] | 30.74 | 0.933 | <0.001 | 35.90 | 0.964 | <0.001 | 28.50 | 0.913 | <0.001 | 33.91 | 0.955 | <0.001 |
| | SANet [30] | 32.04 | 0.948 | <0.001 | 36.63 | 0.968 | <0.001 | 30.01 | 0.935 | <0.001 | 34.62 | 0.960 | <0.001 |
| | McMRSR [31] | 30.92 | 0.932 | <0.001 | 36.94 | 0.971 | <0.001 | 30.65 | 0.940 | <0.001 | 35.13 | 0.965 | <0.001 |
| The proposed | CANM-Net (Ours) | **32.98** | **0.958** | —— | **37.77** | **0.976** | —— | **31.34** | **0.949** | —— | **35.95** | **0.973** | —— |

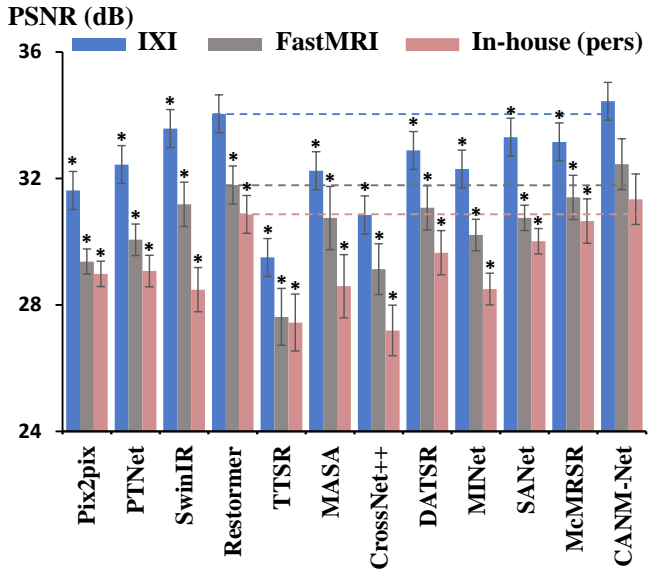

Fig. 7 **Statistical analysis (average and standard deviation) results of the PSNR in the 4× SR task on three datasets**. Our methods have a significantly better performance than others except for the restormer on the IXI dataset. * means p-value < 0.001.

Table IV: Quantitative results of the ablation study.

|  | PSNR | SSIM |
| --- | --- | --- |
| w/o WA | 34.08 | 0.972 |
| w/o CA | 33.53 | 0.968 |
| w/o PS | 33.61 | 0.968 |
| CNN-only | 32.47 | 0.959 |
| w/o FM | 33.71 | 0.969 |
| GFM | 32.66 | 0.958 |
| CANM-Net (Ours) | **34.44** | **0.974** |

A qualitative comparison of the prospective study is shown in **Fig. 7**, where a real-acquired T2WI is selected as the representation. Still, the CANM-Net provides the outputs of the highest quality, clearest anatomic structure, and with the least reconstruction errors, suggesting its potential in the practical acquisition of high-quality MR images.

### C. Ablation Study

#### 1) Ablation Study on Compound Attention:
To study the effectiveness of the compound-attention, we created four models that only have differences in CTLs with the CANM-Net (a) without CABs, named as w/o CA, (b) without WABs, named w/o WA, (c) using WABs and the original channel attention in restormer, without the pyramid-structure in our CABs, named as w/o PS, and (d) only using convolutional layers to extract features, named as CNN-only. We conducted experiments on the IXI dataset and compared their results with the CANM-Net. As shown in **Table 4**, the CNN-only obtains the worst performance among these models. The w/o WA and w/o PS outperforms the w/o CA with the PSNR of 0.55 and 0.08 dB, respectively, which suggests the importance of channel-attention in the image super-resolution. Interestingly, the CANM-Net not only requires less calculation due to the downsampling before self-attention operations, but also has an improvement compared to the w/o PS.

#### 2) Ablation Study on the Feature Matching:
Here we investigate whether different feature matching schemes have an influence on the multi-contrast MRI SR. Similar to the ablation study on self-attention, we also created two models with the following architectures: (a) using global feature matching similar to that of MASA and McMRSR, shorted as GFM and (b) directly concatenating the reference and LR features in the channel dimension, shorted as w/o FM. The quantitative results of the above three networks are summarized in **Table 4**. As can be seen, the performance of these methods drops dramatically compared to CANM-Net with neighborhood-based feature matching, showing the effectiveness of NBFM modules. Moreover, it is observed that the GFM obtains even worse outputs than the w/o FM, suggesting a possible negative influence of the global feature matching.

To further investigate the effect of various feature matching mechanisms, we show the feature maps generated from the 1st and 3rd-level CFE modules of the GFM and CANM-Net, and visualize matching relationships with the highest similarity score between reference and LR features. As shown in **Fig. 8**, the 1st-level feature maps mainly retain low-level characteristics of structural details, and the 3rd-level features have global semantics. At the 1st-level of the GFM, one can find many matching relationships from patches at different positions, which may increase the possibility of mismatching and lead to sub-optimal performance. At the 3rd level of the CANM-Net, the enlarged neighborhood size enables the LR features to receive the non-local contexts of reference features.

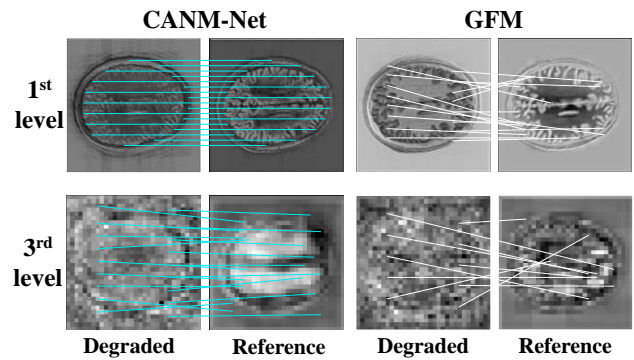

Fig. 8 **Visualization of feature-matching between the degraded (LR) and the reference in the CANM-Net and the GFM**.

### D. Robustness Study

Due to the change of scanning FOV and patient's motion in the practical application, the LR and reference images may have different orientations. The image registration does not always align them well. In this regard, we investigated whether the CANM-Net still has acceptable performance when the reference and LR images are unaligned.

We selected the methods that achieve the top three good results of the 4× SR tasks on the IXI dataset, i.e., restormer,

swinIR, and SANet, in the robustness study. We randomly applied an unalignment with translation and rotation in the range of -4 to 4 pixels and -3 to 3° to each reference image. Quantitative evaluation results are shown in **Table 5**. All the involved methods show decreased performance compared to the well-aligned situations. But CANM-Net still outperforms other baseline methods.

Table V: Quantitative results of robustness study where the reference and LR images are unaligned. The best results are in red.

|  | PSNR | SSIM |
| --- | --- | --- |
| SwinIR [60] | 32.57 | 0.962 |
| Restormer [57] | 33.34 | 0.966 |
| SANet [30] | 31.70 | 0.951 |
| CANM-Net (Ours) | **33.87** | **0.970** |

### E. Efficiency Study

It is important to make a trade-off between performance and efficiency for DL-based methods in practical applications. Here, we created three models with different hyperparameters from the CANM-Net: (a) CANM-tiny, with [2, 2, 12, 2] CTL layers in the $1^{st}$ to $4^{th}$-level CTE modules and 32 channels in the $1^{st}$-level features, (b) CANM-small, [4, 4, 16, 4] CTL layers and 32 channels in the $1^{st}$-level features, and (c) CANM-large, with [6, 6, 16, 6] CTL layers and 96 channels in the $1^{st}$-level features. They are evaluated with the 4× SR tasks on the IXI dataset.

In **Fig. 9**, the horizontal and vertical axes are the number of parameters and PSNR, respectively, and the size of circular areas decreases when the number of float-point operations (FLOPs) increases. Although the CANM-Large outperforms the original CANM-Net with 0.07 dB of PSNR, it has a cost of about two times more parameters and lower efficiency. Compared to baselines, although the CANM-Net requires a larger number of parameters, its superior performance still suggests potential in future applications.

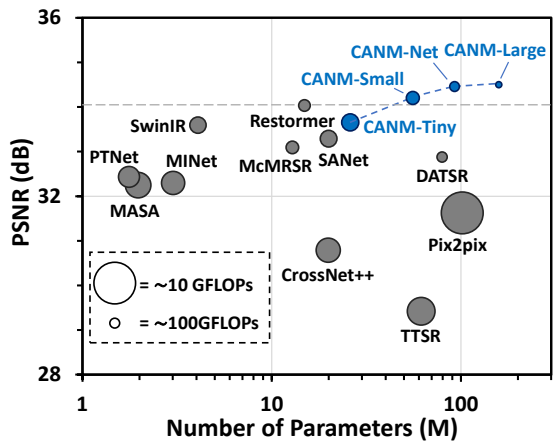

Fig. 9 **The performances of different methods, with their number of model parameters and FLOPs.** The performance is based on the 4× SR tasks on the IXI dataset. A larger size of the circular area represents the corresponding network has lower FLOPs.

## V. DISCUSSION AND CONCLUSION

This paper proposes a novel network architecture with compound attention and neighbor matching (CANM-Net) aiming at the multi-contrast MRI SR tasks. We conducted extensive experiments on the T1WIs and T2WIs from IXI, fastMRI and in-house MR datasets. Our method achieves superior performance to state-of-the-art methods in both prospective and retrospective studies. Moreover, ablation study proves the rationality of our network architecture, and the robustness study shows that our method can still obtain outputs with good quality when the reference and LR images are imperfectly aligned. The results suggest a promising approach for the clinical acquisition of HR multi-contrast MR images.

Several limitations still exist in our work. First, the CANM-Net has large computation complexity, which may limit its applications in hospitals. Second, the parameter tuning is difficult during the network training, due to large number of hyper-parameters. In future research, we will make efforts to reduce the parameters number and the computation cost. Moreover, we may further explore new schemes, e.g., graph neural networks, to further improve the performance.

## REFERENCES

[1] A. E. Chang *et al.*, "Magnetic resonance imaging versus computed tomography in the evaluation of soft-tissue tumors of the extremities," *Ann Surg*, vol. 205, no. 4, pp. 340-348, 1987.

[2] G. Bongartz, T. Vestring, and P. E. Peters, "Magnetic resonance tomography of soft tissue tumors," *Radiologe*, vol. 32, no. 12, pp. 584-590, 1992.

[3] E. Plenge *et al.*, "Super-resolution methods in MRI: can they improve the trade-off between resolution, signal-to-noise ratio, and acquisition time?," *Magn. Reson. Med.*, vol. 68, no. 6, pp. 1983-1993, 2012.

[4] S. A. Mirowitz, "Motion artifact as a pitfall in diagnosis of meniscal tear on gradient reoriented MRI of the knee," *J. Comput. Assist. Tomo.*, vol. 18, no. 2, pp. 279-282, 1994.

[5] R. J. Butlerlewis, W. A. Erdman, H. T. Jayson, B. A. Barker, B. T. Archer, and R. M. Peshock, "Low resolution spin-echo a simple timesaving technique for MRI liver exams," *Magn. Reson. Imag.*, vol. 11, no. 1, pp. 27-33, 1993.

[6] C. Weerasinghe and H. Yan, "Correction of motion artifacts in MRI caused by rotations at constant angular velocity," *Signal Process*, vol. 70, no. 2, pp. 103-114, 1998.

[7] C. Weerasinghe and H. Yan, "An improved algorithm for rotational motion artifact suppression in MRI," *IEEE Trans. Med. Imag.*, vol. 17, no. 2, pp. 310-317, 1998.

[8] H. Greenspan, G. Oz, N. Kiryati, and S. Peled, "Super-resolution in MRI," in *Proc. IEEE Int. Symp. Biom. Imag. (ISBI)*, pp. 943-946, 2002.

[9] I. Despotovic, E. Vansteenkiste, and W. Philips, "Brain volume segmentation in newborn infants using multi-modal MRI with a low inter-slice resolution," *IEEE Eng. Med. Bio.*, pp. 5038-5041, 2010.

[10] C. Granziera *et al.*, "A multi-contrast MRI study of microstructural brain damage in patients with mild cognitive impairment," *Neuroimage-Clin.*, vol. 8, pp. 631-639, 2015.

[11] V. Corona, J. Lellmann, P. Nestor, C. B. Schonlieb, and J. Acosta-Cabronero, "A multi-contrast MRI approach to thalamus segmentation," *Hum. Brain Mapp.*, vol. 41, no. 8, pp. 2104-2120, 1 2020.

[12] E. Karavasilis *et al.*, "Proton density fat suppressed MRI in 3T increases the sensitivity of multiple sclerosis lesion detection in the cervical spinal cord," *Clin. Neuroradiol.*, vol. 29, no. 1, pp. 45-50, 2019.

[13] K. P. Pruessmann, M. Weiger, M. B. Scheidegger, and P. Boesiger, "SENSE: sensitivity encoding for fast MRI," *Magn. Reson. Med.*, vol. 42, no. 5, pp. 952-962, 1999.

[14] F. Huang, J. Akao, S. Vijayakumar, G. R. Duensing, and M. Limkeman, "k-t GRAPPA: A k-space implementation for dynamic MRI with high reduction factor," *Magn. Reson. Med.*, vol. 54, no. 5, pp. 1172-1184, 2005.

[15] S. Bauer, M. Markl, D. Foll, M. Russe, Z. Stankovic, and B. Jung, "K-t GRAPPA accelerated phase contrast MRI: improved assessment of blood flow and 3-directional myocardial motion during breath-hold," *J. Magn. Reson. Imag.*, vol. 38, no. 5, pp. 1054-1062, 2013.

[16] C. H. Chang and J. Ji, "Compressed sensing MRI with multi-channel data using multicore processors," *Magn. Reson. Med.*, vol. 64, no. 4, pp. 1135-1139, 2010.

[17] J. A. Fessler and D. C. Noll, "Iterative image reconstruction in MRI with separate magnitude and phase regularization," in *Proc. IEEE Int. Symp. Biom. Imag. (ISBI)*, pp. 209-212, 2004.

[18] K. T. Block, M. Uecker, and J. Frahm, "Model-based iterative reconstruction for radial fast spin-echo MRI," *IEEE Trans. Med. Imag.*, vol. 28, no. 11, pp. 1759-1769, 2009.

[19] Y. Q. Mohsin, G. Ongie, and M. Jacob, "Accelerated MRI using iterative non-local shrinkage," in *Proc. IEEE Int. Conf. Eng. in Med. and Biol. Soc. (EMBC)*, pp. 1545-1548, 2014.

[20] J. S. Li, Q. G. Liu, and J. Zhao, "Self-prior image-guided MRI reconstruction with dictionary learning," *Med. Phys.*, vol. 46, no. 2, pp. 517-527, 2019.

[21] G. Cruz *et al.*, "Low-rank motion correction for accelerated free-breathing first-pass myocardial perfusion imaging," *Magn. Reson. Med.*, vol. 90, no. 1, pp. 64-78, 2023.

[22] J. Z. Huang, C. Chen, and L. Axel, "Fast multi-contrast MRI reconstruction," *Magn. Reson. Imag.*, vol. 32, no. 10, pp. 1344-1352, 2014.

[23] E. H. Gong, F. Huang, K. Ying, W. C. Wu, S. Wang, and C. Yuan, "PROMISE: parallel-imaging and compressed-sensing reconstruction of multi-contrast imaging using sharable information," *Magn. Reson. Med.*, vol. 73, no. 2, pp. 523, 2015.

[24] Z. Z. Chen *et al.*, "Joint reconstruction of multi-contrast images and multi-channel coil sensitivities," *Appl. Magn. Reson.*, vol. 48, no. 9, pp. 955-969, 2017.

[25] K. Zeng, H. Zheng, C. B. Cai, Y. Yang, K. H. Zhang, and Z. Chen, "Simultaneous single- and multi-contrast super-resolution for brain MRI images based on a convolutional neural network," *Comput. Biol. Med.*, vol. 99, pp. 133-141, 1 2018.

[26] D. Polak *et al.*, "Joint multi-contrast variational network reconstruction (jVN) with application to rapid 2D and 3D imaging," *Magn. Reson. Med.*, vol. 84, no. 3, pp. 1456-1469, 2020.

[27] B. J. Zou, Z. X. Ji, C. Z. Zhu, Y. L. Dai, W. S. Zhang, and X. Y. Kui, "Multi-scale deformable transformer for multi-contrast knee MRI super-resolution," *Biomed. Signal. Proces.*, vol. 79, 2023.

[28] Q. Lyu *et al.*, "Multi-contrast super-resolution MRI through a progressive network," *IEEE Trans. Med. Imag.*, vol. 39, no. 9, pp. 2738-2749, 2020.

[29] C. M. Feng, H. Z. Fu, S. H. Yuan, and Y. Xu, "Multi-contrast MRI super-resolution via a multi-stage integration network," in *Proc. Int. Conf. Med. Image Comput. Comput.-Assisted Intervent. (MICCAI)*, pp. 140-149, 2021.

[30] C. M. Feng *et al.*, "Exploring separable attention for multi-contrast MR image super-resolution," *arxiv preprint*, arXiv:2109.01664.

[31] G. Y. Li *et al.*, "Transformer-empowered multi-scale contextual matching and aggregation for multi-contrast MR image super-resolution," in *Proc. IEEE Conf. Comput. Vis. Pattern Recognit. (CVPR)*, pp. 20604-20613, 2022.

[32] C. M. Feng, Yan, L., Cheng, G., Xu, Y., Hu, Y., Shao, L., Fu, H., "Multi-modal transformer for accelerated MR imaging," *IEEE Trans. Med. Imag.*, 2022.

[33] A. Vaswani *et al.*, "Attention is all you need," in *Proc. Conf. Neural Inf. Process. Syst.*, vol. 30, 2017.

[34] L. Y. Lu, W. B. Li, X. Tao, J. B. Lu, and J. Y. Jia, "MASA-SR: matching acceleration and spatial adaptation for reference-based image super-resolution," in *Proc. IEEE Conf. Comput. Vis. Pattern Recognit. (CVPR)*, pp. 6364-6373, 2021.

[35] "IXI Dataset," http://brain-development.org/ixi-dataset/

[36] F. Knoll *et al.*, "fastMRI: A publicly available raw k-space and DICOM dataset of knee images for accelerated MR image reconstruction using machine learning," *Radiol-Artif. Intell.*, vol. 2, no. 1, 2020. https://fastmri.med.nyu.edu/

[37] N. Karani, I. Zhang, C. Tanner, and E. Konukoglu, "An image interpolation approach for acquisition time reduction in navigator-based 4D MRI," *Med. Image Anal.*, vol. 54, pp. 20-29, 2019.

[38] J. Joo, K. H. Jin, H. Gupta, J. Yerly, M. Stuber, and M. Unser, "Time-dependent deep image prior for dynamic MRI," *IEEE Trans. Med. Imag.*, vol. 40, no. 12, pp. 3337-3348, 2021.

[39] X. G. Li, Y. M. Sun, Z. P. Liu, Y. L. Yang, and C. Y. Miao, "Dense residual network for X-ray images super-resolution," in *Proc. IEEE 3rd International Conference on Cloud Computing and Big Data Analysis (ICCCBDA)*, pp. 336-340, 2018.

[40] S. Rochester, D. English, I. Lacey, K. Munechika, and V. V. Yashchuk, "Towards super-resolution interference microscopy metrology of X-ray variable-line-spacing diffraction gratings: Recent developments," *Proc. Spie.*, vol. 12240, 2022.

[41] Y. L. Zhang, Y. P. Tian, Y. Kong, B. N. Zhong, and Y. Fu, "Residual dense network for image super-resolution," in *Proc. IEEE Conf. Comput. Vis. Pattern Recognit. (CVPR)*, pp. 2472-2481, 2018.

[42] Z. Y. An, J. Y. Zhang, Z. Y. Sheng, X. H. Er, and J. J. Lv, "RBDN: residual bottleneck dense network for image super-resolution," *IEEE Access*, vol. 9, pp. 103440-103451, 2021.

[43] S. Zhou, L. Yu, and M. Jin, "Texture transformer super-resolution (TTSR) for patient CT images," *Med. Phys.*, vol. 49, no. 6, pp. E591-E591, 2022.

[44] R. Z. Shilling, T. Q. Robbie, T. Bailloeul, K. Mewes, R. M. Mersereau, and M. E. Brummer, "A super-resolution framework for 3-D high-resolution and high-contrast imaging using 2-D multislice MRI," *IEEE Trans. Med. Imag.*, vol. 28, no. 5, pp. 633-644, 2009.

[45] C. M. Feng, Y. L. Yan, H. Z. Fu, L. Chen, and Y. Xu, "Task transformer network for joint MRI reconstruction and super-resolution," in *Proc. Int. Conf. Med. Image Comput. Comput.-Assisted Intervent. (MICCAI)*, vol. 12906, pp. 307-317, 2021.

[46] Y. L. Zhang, K. Li, K. P. Li, and Y. Fu, "MR image super-resolution with squeeze and excitation reasoning attention network," in *Proc. IEEE Conf. Comput. Vis. Pattern Recognit. (CVPR)*, pp. 13420-13429, 2021.

[47] J. S. Hong *et al.*, "Acceleration of magnetic resonance fingerprinting reconstruction using denoising and self-attention pyramidal convolutional neural network," *Sensors-Basel.*, vol. 22, no. 3, 2022.

[48] L. Xiang *et al.*, "Deep-learning-based multi-modal fusion for fast MR reconstruction," *IEEE Trans. Bio-Med. Eng.*, vol. 66, no. 7, pp. 2105-2114, 2019.

[49] Y. Dai, Y. F. Gao, and F. Y. Liu, "TransMed: transformers advance multi-modal medical image classification," *Diagnostics*, vol. 11, no. 8, 2021.

[50] H. T. Zheng, M. Q. Ji, H. Q. Wang, Y. B. Liu, and L. Fang, "CrossNet: an end-to-end reference-based super resolution network using cross-scale warping," in *Proc. Euro. Conf. on Comput. Vis. (ECCV)*, pp. 87-104, 2018.

[51] F. Z. Yang, H. Yang, J. L. Fu, H. T. Lu, and B. N. Guo, "Learning texture transformer network for image super-resolution," in *Proc. IEEE Conf. Comput. Vis. Pattern Recognit. (CVPR)*, pp. 5790-5799, 2020.

[52] Y. X. Wu *et al.*, "Multi-contrast MRI registration of carotid arteries based on cross-sectional images and lumen boundaries," in *Medical Imaging 2017: Image Processing*, vol. 10133, 2017.

[53] X. Huang and S. Belongie, "Arbitrary style transfer in real-time with adaptive instance normalization," in *Proc. IEEE Conf. Comput. Vis. Pattern Recognit. (CVPR)*, pp. 1510-1519, 2017.

[54] A. Dosovitskiy *et al.*, "An image is worth 16×16 words: transformers for image recognition at scale," *arxiv preprint*, arXiv:2010.11929., 2020.

[55] J. B. Zhu, M. F. Ge, Z. M. Chang, and W. F. Dong, "GCCSwin-UNet: global context and cross-shaped windows vision transformer network for polyp segmentation," *Processes*, vol. 11, no. 4, 2023.

[56] Z. Liu *et al.*, "Swin transformer: hierarchical vision transformer using shifted windows," in *Proc. IEEE Int. Conf. Comput. Vis. (ICCV)*, pp. 9992-10002, 2021.

[57] S. W. Zamir, A. Arora, S. Khan, M. Hayat, F. S. Khan, and M. H. Yang, "Restormer: efficient transformer for high-resolution image restoration," in *Proc. IEEE Conf. Comput. Vis. Pattern Recognit. (CVPR)*, pp. 5718-5729, 2022.

[58] P. Isola, J. Y. Zhu, T. H. Zhou, and A. A. Efros, "Image-to-image translation with conditional adversarial networks," in *Proc. IEEE Conf. Comput. Vis. Pattern Recognit. (CVPR)*, pp. 5967-5976, 2017.

[59] X. Z. Zhang *et al.*, "PTNet3D: A 3D high-resolution longitudinal infant brain MRI synthesizer based on transformers," *IEEE Trans. Med. Imag.*, vol. 41, no. 10, pp. 2925-2940, 2022.

[60] J. Y. Liang, J. Z. Cao, G. L. Sun, K. Zhang, L. Van Gool, and R. Timofte, "SwinIR: image restoration using swin transformer," in *Proc. IEEE Int. Conf. Comput. Vis. (ICCV)*, pp. 1833-1844, 2021.

[61] Y. Tan *et al.*, "CrossNet plus plus : cross-scale large-parallax warping for reference-based super-resolution," *IEEE Trans. Pattern Anal. Mach. Intell.*, vol. 43, no. 12, pp. 4291-4305, 1 2021.

[62] J. Z. Cao *et al.*, "Reference-based image super-resolution with deformable attention transformer," in *Proc. Euro. Conf. on Comput. Vis. (ECCV)*, pp. 325-342, 2022.